\documentclass[seceq,preprint]{ptptex}

\usepackage{graphicx}

\preprintnumber[3cm]{OU-HET 523\\KUNS-1965\\YITP-05-15\\OIQP-05-03}

\title{Cyclic Universe \`{a} la string theory}

\author{Yoshinobu \textsc{Habara}$^{a)}$\footnote{Adress after 1 April 2005, Department of Physics, Kyoto University, Kyoto 606-8502, Japan.}, Hikaru \textsc{Kawai}$^{b)c)}$ and Masao \textsc{Ninomiya}$^{d)}$\footnote{Also working at Okayama Institute for Quantum Physics.}}

\inst{$^{a)}$Department of Physics, Osaka University, Osaka 560-0043, Japan\\
$^{b)}$Department of Physics, Kyoto University, Kyoto 606-8502, Japan\\
$^{c)}$Theoretical Physics Laboratory, RIKEN, Wako 351-0198, Japan\\
$^{d)}$Yukawa Institute for Theoretical Physics, Kyoto University,\\Kyoto 606-8502, Japan}

\begin{document}

\maketitle

\section{Introduction}

Recently, the observations in cosmology have been exceedingly accurate such as WMAP~\cite{rf:1,rf:2}, and pre-big bang physics has started to be investigated quantitatively. So far observational results by COBE etc. support the inflation scenario: This scenario is believed to solve the flatness and horizon problem in a simple manner. However, in view of the field theory model, the form of the potential of the fictious scalar particle called ``inflaton" is very unnatural and peculiar. That is yet to be investigated.

In this paper, We shall present a model of the ``cyclic universe"~\cite{rf:3,rf:4,rf:5} that can be constructed only by assuming a  minimal set of properties of string theory. We clarify our viewpoint of the cyclic universe and show some attempts to mateliarize the idea as field theoretical manner.

The contents are the following: In section 2, We briefly explain the basic notion of the string theory needed for understanding the cyclic universe and some requirements to the cosmological model. In section 3, we trace back in time from the present universe to the moment of birth of our universe. Here, in our view, the size of the universe around the big bang moment is the order of the strig scale. Thus we can further go back and reaches the big crunch. The cyclic universe scenario consists of repetition of the big bangs and big crunches. In section 4, we explain that, in the process of this repetiton, the universe stores entropy and finally gains the present day's enormously huge entropy. In section 5, we present various attempts in order to realize cyclic universe picture and point out some problems to overcome.

\section{Basic notion of the string theory}

As is well known, in the string theory there exists the minimal string length $l_s$ of the order of 10 to minus thirty-three centi-meters. The novel feature is that the particles and their masses appear from the oscillation of the string which is internal degrees of freedom. From this fact, it is wellknown that there exists an upper limit of temperature. This limiting temperature is derived by considering the partition function, 

\begin{align}
	Z(T)\propto \int_{0}^{\infty}dm\> e^{ml_s}e^{-m/T}.
\end{align}

\noindent This equation means the density of the number of states $\rho_{\text{num}} (m)$ in the mass spectrum of the string is proportional to $e^{ml_s}$, 

\begin{align}
	\rho_{\text{num}} (m)\propto e^{ml_s}.
\end{align}

\noindent In order to avoid the divergence of the partition function $Z(T)$, the temperature must have the upper limit called Hagedorn temperature $T_H$, 

\begin{align}
	T<m_s=\frac{1}{l_s}\equiv T_H.
\end{align}

\noindent Implication of this fact is that as the temperature approaches to the upper limit $T_H$, the energy of string flows, not into momentum but into oscillations which is nothing but the masses, and thus the higher excited states with higher masses are created.

These peculiar properties of the string theory indicate that, when going back along the time from present universe, the size of the universe cannot be smaller than that of the srting length $l_s$, and the temperature must be less than the Hagedorn one $T_H$. Furthermore, as the temperature approaches about $T_H$, the correction to the Einstein equation from the string's higher order effects becomes bigger and bigger. In this way, the solution to the Einstein equation may not describe the collapse of the universe.

Let us explain in more detail the above statement. It is known that the Hawking temperature $T_{HG}$ of the de Sitter expanding universe is proportional to the Hubble constant: 

\begin{align}
	T_{HG}\propto H=\frac{\dot{a}(t)}{a(t)}.
\end{align}

\noindent Here $a(t)$ denotes the radius of the universe. From this fact, it is natural to assume that there should be an upper limit of the expansion rate of the universe: 

\begin{align}
	\frac{\dot{a}(t)}{a(t)}<m_s=T_H.
\end{align}

\begin{figure}[htbp]
\begin{center}
\resizebox{!}{70mm}{
\includegraphics{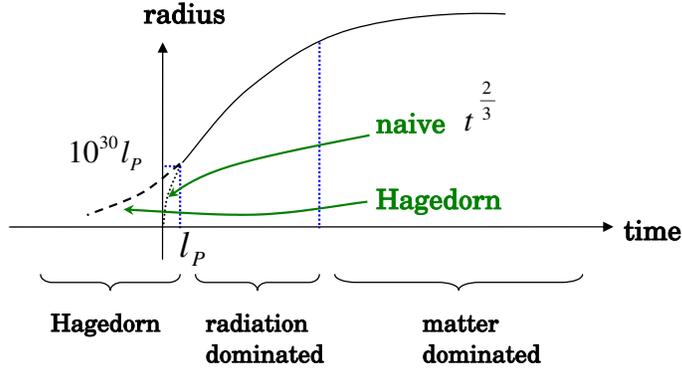}}
\end{center}
\caption{\footnotesize{Era of the Hagedorn universe}}
\label{fig1}
\end{figure}

\noindent Then, the universe with the temperature about the upper limit $T_H$ may expand in time with upper limit rate $m_s$, 

\begin{align}
	\frac{\dot{a}(t)}{a(t)}=m_s \> \Longrightarrow \> a(t)\propto e^{m_st}.
\end{align}

\noindent so that the universe turns out to expand exponentially. We may call this limiting universe with exponential expansion ``Hagedorn universe". Let us summarize the scenario so far which is presented in Fig. \ref{fig1}.

\section{Cyclic Universe}

Then, as we trace back in time, the Hagedorn universe shrinks exponentially. But in the string theory, there exists the minimal length $l_s$, so that in our cyclic universe, the size cannot be smaller than $l_s$. If we invoke the fundamental properties of the string theory, called T-duality, we may obtain that the size of the universe will expand after reaching the minimal size $l_s$. In more detail, suppose that the size of the universe evolves exponentially: 

\begin{align}
	a(t)=l_se^{m_st}.
\end{align}

\noindent If we impose the T-duality on $a(t)$ in the following manner:  

\begin{align}
	a(t)\> \longleftrightarrow \> \frac{l_s^2}{a(t)},
\end{align}

\noindent the size of the universe, after reaching the minimal value $l_s$, starts to again expand exponentially when tracing back in time,

\begin{align}
	a(t)=\frac{l_s^2}{l_se^{m_st}}=l_se^{-m_st}.
\end{align}

\begin{figure}[htbp]
\begin{center}
\resizebox{!}{50mm}{
\includegraphics{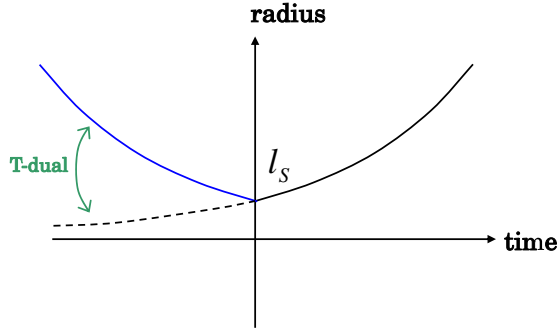}}
\end{center}
\caption{\footnotesize{Bounce via T-duality}}
\label{fig2}
\end{figure}

\noindent See Fig. \ref{fig2} representing the bounce. 

\begin{figure}[ht]
\begin{center}
\resizebox{!}{50mm}{
\includegraphics{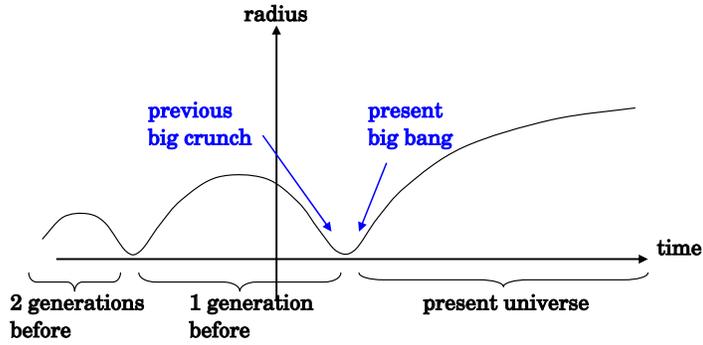}}
\end{center}
\caption{\footnotesize{Cyclic Universe}}
\label{fig3}
\end{figure}

From the above argument, when tracing back the universe in the past direction before the big bang of our universe, there was the big crunch one generation before. In this way, we can depict the cyclic universe as Fig. \ref{fig3}.

\section{Entropy production}

Now We adress our attension to the mechanism how the entropy is produced at the moment of the big bang and big crunch. In the present universe, we may estimate numerical value of the entropy given by 

\begin{align}
	S \simeq T^3V.
\end{align}

\noindent In the beginning of the radiation dominated era, the temperature is given by the order of the string scale $m_s$, and the volume $10^{30}l_s$, 

\begin{align}
	& \text{temperature:}\quad T\sim m_s\sim \frac{1}{l_s}, \\
	& \text{volume:}\quad V\sim (10^{30}l_s)^3.
\end{align}

\noindent Then we may estimate the present entropy as 
\begin{align}
	S\sim T^3V\sim O(10^{90}).
\end{align}

\noindent We would like to interpret this tremendously big value of the entropy is achieved through a series of the entropy productions by big bangs and big crunches. If we assume that the size of the first generation uiverse is of order of 1, string scale, similarly the entropy is also of the same order.

It turns out that the big bang and big crunch can be considered in the same manner as far as the entropy production is concerned, so that we may investigate the entropy production for big crunch.

First of all, we notice that in the radiation dominated era, the radiation adiabatically changes in time, and thus there is no entropy production. Also in the Hagedorn era, since the universe is filled by the highly excited strings with extremely high density, the expansion rate is of the order of string scale $m_s$, while the energy density is enormously big compared to $m_s$, and thus the change becomes adiabatic. From this consideration, the entropy can only increase when the universe transits from the radiation dominated era into the Hagedorn era. The transition is realized through the string scattering, i.e. the massless particles into the massive ones.

\begin{figure}[ht]
\begin{center}
\resizebox{!}{40mm}{
\includegraphics{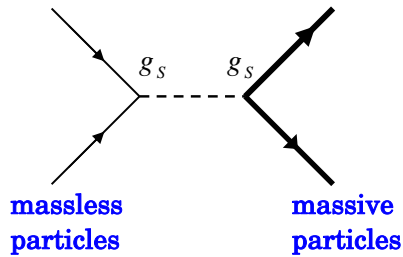}}
\end{center}
\caption{\footnotesize{Massive particle production from the massless ones scattering}}
\label{fig4}
\end{figure}

\noindent In this era, since the universe shrinks exponentially, we may assume the transition or relaxation time as the order of $l_s \sim 3l_s$. 

\begin{figure}[ht]
\begin{center}
\resizebox{!}{50mm}{
\includegraphics{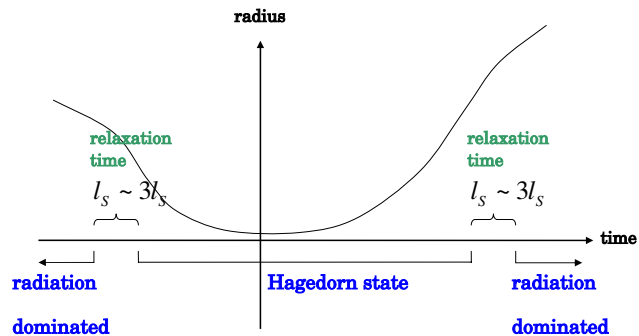}}
\end{center}
\caption{\footnotesize{Periods of entropy-production}}
\label{fig5}
\end{figure}

During this period, the entropy produced is estimated as follows. Suppose that the Hamiltonian for a mode with the oscillation $\omega$ is given by the following $H_{\omega}$,

\begin{align}
	H_{\omega}=\omega \left( \frac{1}{2}p^2+\frac{1}{2}q^2\right).
\end{align}

\noindent which is nothing but that of the harmonic oscillator. The energy of this mode is identified to the temperature $m_s$ because of the equipartition law. Thus the entropy is obtained by deviding by $2\pi$ the area of the ellipse with the energy lower than $m_s$ in the phase space, and then taking logarithm. If we assume that the oscillation number is also cut by a maximal value $m_s$, the average entropy of each mode $S_{\text{before}}$ is obtained,  

\begin{align}
	S_{\text{before}}=\frac{\displaystyle \int_{\omega <m_s}
	\frac{d^3\omega}{(2\pi)^3}\log \left(\frac{m_s}{\omega}\right)}
	{\displaystyle \int_{\omega <m_s}\frac{d^3\omega}{(2\pi)^3}1}
	=\frac{1}{3}.
\end{align}

Next, we estimate the change of the entropy of each mode, when the size of the universe varies. In fact, during the relaxation time $\tau$ of the scattering equals to $l_s \sim 3l_s$, the universe gets smaller by a factor $s$, 

\begin{align}
	s=e^{-m_s\tau}.
\end{align}

\noindent Now, suppose that the change of the size of the universe is much slower than that of the oscillation, each mode will adiabatically develop in time, and thus the ellipse in the phase space changes by keeping its area. 

\begin{figure}[ht]
\begin{center}
\resizebox{!}{50mm}{
\includegraphics{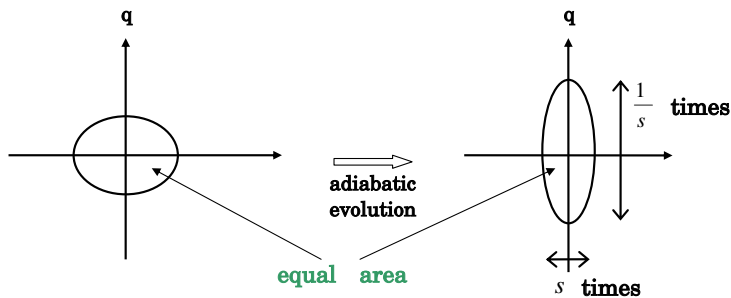}}
\end{center}
\caption{\footnotesize{Adiabatic evolution}}
\label{fig6}
\end{figure}

\noindent As is depicted in Fig. \ref{fig6}, the change is given by multiplying $\frac{1}{s}$ in q-axis direction and $s$ in p-axis direction, so that the area is unchanged and thus the entropy is also unchanged. In summary, the entropy, after this adiabatic evolution, $S_{\text{adiabatic}}$ keeps its value.

\begin{align}
	S_{\text{adiabatic}}=S_{\text{before}}=\frac{1}{3}
\end{align}

\begin{figure}[ht]
\begin{center}
\resizebox{!}{50mm}{
\includegraphics{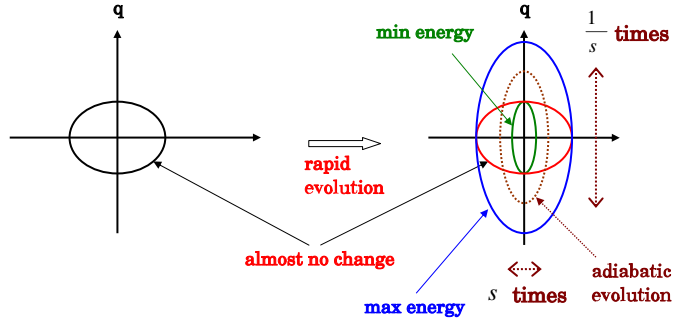}}
\end{center}
\caption{\footnotesize{Non-adiabatic evolution}}
\label{fig7}
\end{figure}

However, in reality, the change of the size of the universe occurs very rapidly, so that, before each mode develops in time, the size of the universe get multiplied by the factor $s$. Thus, as is shown in Fig. \ref{fig7}, the energy fluctuates between $s^2$ and $\frac{1}{s^2}$ multiplications.

\noindent Therefore, the maximal value of the area of equi-energy surface is $\frac{1}{s^2}$ times the original area, and we gain the entropy non-adiabatically by taking the logarithm of $\frac{1}{s^2}$, 

\begin{align}
	S_{\text{non-adiabatic}}=S_{\text{before}}+\log s^{-2}
	=\frac{1}{3}+\log (e^{-m_s\tau})^{-2}=\frac{1}{3}+2m_s\tau.
\end{align}

\noindent From this calculation, we may say that each mode produces the entropy by an amount of $2m_s\tau$. That is to say, for the big crunch, the entropy is multiplied by $1+6m_s\tau$ to become, 

\begin{align}
	\frac{S_{\text{non-adiabatic}}}{S_{\text{adiabatic}}}=
	\frac{\frac{1}{3}+2m_s\tau}{\frac{1}{3}}=1+6m_s\tau.
\end{align}

\noindent By using the relaxaion time $\tau$ is about $l_s \sim 3l_s$, we may rewrite this ratio as 

\begin{align}
	\frac{S_{\text{non-adiabatic}}}{S_{\text{adiabatic}}}=7\sim 20.
\end{align}

\noindent Thus we conclude that the entropy is multiplied about $7 \sim 20$ times compared to the adiabatic evolution. Since we consider that the entropy increases in this mechanism, our present universe has the entropy 

\begin{align}
	S_{\text{now}}\sim O(10^{90}).
\end{align}

\noindent From this entropy value, we may conclude that the present universe is 35th or 45th generation from the birth of the universe.

\section{Model building and conclusions}

In this last section, We would like to present some attempts for constructing field theoretical models that reveals properties of the cyclic universe. 

As was explained before, in the Hagedorn era, in which the temperature takes the maximal value, the universe is filled by the higher excited strings with exceedingly high density. Such universe may be described by higher derivative gratitational theories. In fact, we consider the Einstein-Hilbert action with a correction including the Ricci scalar $R$, 

\begin{align}
	S=\int d^4x\> \sqrt{-g}\> P(R)
	+\int d^4x\> \sqrt{-g}\> \rho (x)Q(R).
\end{align}

\noindent Here, $\rho (x)$ denotes the matter density. Furthermore the massive string can be viewed as an ordinary scalar field, so that we may write the matter density as 

\begin{align}
	\rho (x)=\frac{M_d}{a^3(t)}.
\end{align}

\noindent where $M_d$ is the total mass in the universe. The explicit forms for the Lagrangian we tried as the corrections from the higher excited states of the strings are the followings, where $P(Q)$ denotes the pure gravity term and $Q(R)$ the correction to matter.

\begin{align}
	P(R), Q(R)=
	\left\{ \begin{array}{ll}
	-R+zR^2, & 
	\text{\small (or in general, including higher order polynomial)}\\
	\displaystyle -\frac{R}{1+R}, & \text{\small (rational function)} \\
	1-\sqrt{1+2R}, & \text{\small (irrational function)}
	\end{array} \right.
\end{align}

\noindent Here we introduced a parameter $z$ for later convenience.

Another type of variation of the action is that of the Born-Infeld type given by 

\begin{align}
	S=\int d^4x\> \left\{ -\sqrt{det(R_{\mu \nu}-g_{\mu \nu})}
	+\sqrt{det(-g_{\mu \nu})}\right\} \quad \text{(Born-Infeld type)}.
\end{align}

\noindent We then use as the space-time metric the Freedman-Robertson-Walker metric: 

\begin{align}
	ds^2=-dt^2+a^2(t)d\Omega_3^2.
\end{align}

\begin{figure}[ht]
\begin{center}
\rotatebox{270}{{
\includegraphics[height=7.5cm,keepaspectratio]{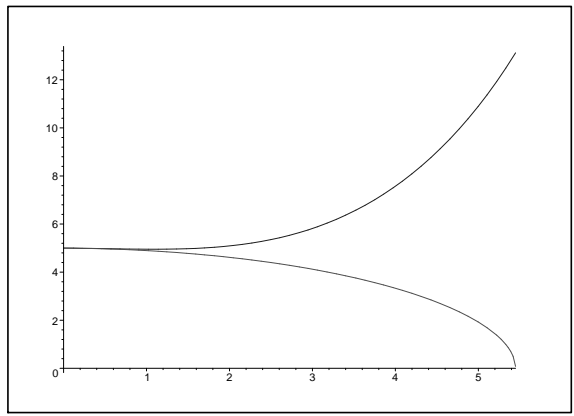}}}
\end{center}
\caption{\footnotesize{$R^2$ gravity model: The convex line is the result of the Einstein gravity, while the concave line is that of the higher derivative theories.}}
\label{fig8}
\end{figure}

The numerical investigations have been carried out for these models. The typical result is depicted in Fig. \ref{fig8}. Here, the vertical axis represents the scale factor $a(t)$ and the horizontal one time $t$, and the convex line is the ordinary Einstein gravity case, while the concave line shows the gravity with higher order derivative terms.

\noindent From this result, we might think that the solution of Einstein thoery collapses, while the higher derivative theories might look very promising because the line is bounded. However, there is a severe problem that the line goes to infinity after bounded.

In order to clarify the problem, we consider the linearization by utilizing the $R^2$ gravity. We decompose $a(t)$ into two parts: one is the solution of the Einstein equation, $a_E$, and the other is a correction $q$ as follows, 

\begin{align}
	a(t)=a_E(t)+zq(t).
\end{align}

\noindent By considering the zeroth and first order in $z$ are given by these equations of motion.

\begin{align}
	& O(z^0):\quad 6\dot{a}_E^2(t)+6k=\frac{M_d}{a_E(t)} \quad 
	(\text{Einstein eq.}) \\
	& O(z^1):\quad \dot{q}(t)
	=\frac{1}{2}\frac{M_d}{a_E(t)\sqrt{6a_E(t)(M_d-6ka_E(t))}}q(t) 
	\nonumber \\
	& \qquad \qquad \qquad \qquad \qquad +\frac{1}{8}\frac{-72ka_E(t)+9M_d}
	{a_E^3(t)\sqrt{6a_E(t)(M_d-6ka_E(t))}} \\
	& \qquad \qquad \qquad \sim \frac{1}{2\sqrt{6}}
	\sqrt{\frac{M_d}{a_E^3(t)}}q(t)+\cdots, 
	\qquad a_E(t)\sim 0 \\
	& \qquad \qquad \qquad \sim \frac{1}{2\sqrt{6}}\sqrt{\rho (x)}
	q(t)+\cdots
\end{align}

\noindent Needless to say that the zeroth order equation is nothing but the Einstein equation, while the first order one is linear in $q$. Now, in the equation for $q(t)$, the coefficient is precisely square root of the matter density $\rho (x)$.

\begin{align}
	\rho (x)=\frac{M_d}{a_E^3(t)}
\end{align}

\noindent Therefore when the matter field is singular at $a(t)$ equals zero, $\dot{q}(t)$ diverges. 

We also tried to supress this divergence by introducing a friction. For example, we added the friction term to the equation for $R^2$ gravity.

\begin{align}
	\text{(terms from $R^2$)}+6\dot{a}_E^2(t)+6k=\frac{M_d}{a_E(t)}
	+\text{(friction term)}
\end{align}

\noindent Here, for example, the friction term is taken as those proportional to the Hubble constant. Including the friction term indicates that it makes the accelaration of the space-time flow into the string oscillation, and finally makes the system finite. However this idea does not work, because, if we utilize the classical gravitational theories including the higher derivative terms, only one bound can be realized, but repetition of the big bang and big crunch cannot be realized because the scale of the universe $a(t)$ goes to infinity.

\vspace{0.5cm}

\noindent \underline{ Conclusions }

\vspace{0.25cm}

If the string theory is the theory of everything, the size of the universe is bounded from below, and at the same time the energy and temperature have upper bounds. Then we may derive naturally a picture of cyclic universe from these basic properties of the string theory. And the model building was considered only in classical methods by adding the stringy effects, the higher excited states of the strings. One of our conclusions is that by such method we may not be able to construct the model which shows many times bounds. The difficulty seems to come from singularity of the matter potential at the origin.

\section*{Acknowledgements}
Y.H. is supported by The 21st century COE Program ``Towards a New Basic Science; Depth and Synthesis", of the Department of Physics, Graduate School of Science, Osaka University. This work is supported by Grants-in-Aid for Scientific Research on Priority Areas, Number of Areas 763, ``Dynamics of strings and Fields", from the Ministry of Education of Culture, Sports, Science and Technology, Japan.

%


\begin{thebibliography}{99}
  
\bibitem{rf:1}E. Komatsu et al., arXiv:astro-ph/0302207; astro-ph/0302209; astro-ph/0302225

\bibitem{rf:2}Y. Suto, http://www-utap.phys.s.u-tokyo.ac.jp/~suto/index.html

\bibitem{rf:3}G. Lema$\hat{\text{i}}$tre, C. R. Acad. Sci. (Paris) 196 (1933) 903; Ann. Soc. Sci. Brussels A53 (1933) 85

\bibitem{rf:4}M. Fukuma, H. Kawai and M. Ninomiya, IJMPA Vol.19, No.26 (2004) 4367-4385

\bibitem{rf:5}P.J. Steinherdt and N. Turok, arXiv:hep-th/0111030; Phys.Rev. D65, 126003 (2002) p443


\end{thebibliography}
\end{document}